\documentclass[prl]{revtex4}
\usepackage[dvips]{graphicx}

\usepackage{amssymb}

\begin{document}

\title{Gluon Quasiparticles and the Polyakov Loop}

\author{Peter N.\ Meisinger}
\author{Michael C.\ Ogilvie}
\affiliation{
Dept.\ of Physics
Washington University
St.\ Louis, MO 63130}

\author{Travis R.\ Miller}
\affiliation{
Dept.\ of Physics
Indiana University
Bloomington, IN 47405}

\begin{abstract}
A synthesis of Polyakov loop models of the deconfinement transition and
quasiparticle models of gluon plasma thermodynamics leads to a class of
models in which gluon quasiparticles move in a non-trivial 
Polyakov loop background. These
models are successful candidates for explaining both critical behavior and
the equation of state for the $SU(3)$ gauge theory at temperatures above the
deconfinement temperature $T_{c}$. Polyakov loops effects are most important
at intermediate temperatures from $T_{c}$ up to roughly $2.5T_{c}$, while
quasiparticle mass effects provide the dominant correction to blackbody
behavior at higher temperatures.
\end{abstract}

\maketitle

The equation of state is one point where theory, experiment, and lattice
calculation all provide valuable contributions to our understanding of
finite temperature QCD. Perturbation theory predicts the equation of state
at high temperatures, with interactions providing corrections to the
blackbody expression. Experiment typically uses phenomenological
equations of state as input for modelling and analysis of data. Lattice
gauge theory simulations provide a body of results on the critical
behavior and thermodynamics of QCD and related theories at finite
temperature. These results are among the most important that lattice gauge
theory simulations have provided, giving important guidance to both theory
and experiment. Analytical work on the thermodynamics of the quark-gluon
plasma at zero chemical potential is now benchmarked against results from
lattice simulations, but theoretical work derived solely from QCD has had
difficulty explaining lattice results at intermediate temperatures, just
above the temperature where the quark-gluon plasma forms. Perturbation
theory for the pressure has been calculated to $O\left( g^{5}\right) $
\cite{Zhai:1995ac,Braaten:1995ju},
but the series is not well-behaved, and the prospects for extracting the $%
O\left( g^{6}\right) $ contribution from the infinite set of diagrams
required are not good\cite{Kapusta:tk,Linde:ts}.
Closely related approaches are dimensional reduction
\cite{Kajantie:2000iz,Kajantie:2002wa}
and hard thermal loop (HTL) resummation
\cite{Blaizot:1999ip,Blaizot:1999ap,Blaizot:2000fc,Peshier:2000hx,Andersen:1999fw}.
The former can extract the $O\left(
g^{6}\ln g\right) $ contribution to the equation of state, but not the full $%
O\left( g^{6}\right) \,$behavior; the fit to the lattice data is extremely
sensitive to the choice of scale. The latter appears to be capable of
fitting lattice results well at high temperatures, but again falls short at
intermediate temperatures. More significantly, perturbation theory and its
variants cannot account for confinement and chiral symmetry breaking at low
temperatures, as well as the rather wide range of phase transitions and other
critical behavior observed in lattice simulations of QCD and related models.

Phenomenological models of deconfinement and chiral symmetry
restoration at finite temperature have been succesful at explaining much of
the phase structure of QCD and related models. Linear sigma
models \cite{Goloviznin:1992ws,Greiner:wv}
and Nambu-Jona Lasinio models 
\cite{Klevansky:qe,Schwarz:1999dj}
have been used extensively to study
chiral symmetry as quark properties, \textit{e.g.}, the number of light
quarks, are varied. Because these models do not include the gluonic
degrees of freedom, they are incapable of completely describing the
quark-gluon plasma. Models based on the use of the Polyakov loop as
the order parameter for confinement play a somewhat parallel role in
modeling the deconfinement transition in pure gauge theories
\cite{Pisarski:2000eq,Dumitru:2000in,Meisinger:2001cq}.
However, in
their simplest form such models have as their principal success the
prediction of the order of the phase transition for $SU(2)$ and $SU(3)$, and
do not make detailed predictions of thermodynamic behavior.

A third theoretical approach to QCD thermodynamics combines perturbative
results with a simple physical picture. Quasiparticle models have been quite
successful in fitting lattice results over a large range of temperatures 
\cite{Peshier:1995ty,Levai:1997yx,Schneider:2001nf}.
In the case of pure $SU(3)$ thermodynamics, 
some models fit lattice results at
all temperatures above the deconfinement transition. All such models treat
the quark-gluon plasma as a gas of independent
quasiparticles with temperature-dependent effective masses accounting for
much of the deviation from blackbody thermodynamics. However, many
quasiparticle must introduce \textit{ad hoc} mechanisms to turn off gluonic
degrees of freedom at low temperatures.

Here we focus on pure $SU(3)$ gauge theory, particularly the deconfinement
phase transition and the thermodynamics of the gluon plasma. We show that
the thermodynamic behavior above the deconfinement temperature $T_{c}$ can
be largely explained in terms of quasiparticle gluons moving in the presence
of a non-trivial background Polyakov loop. As will be discussed below, the
Polyakov loop naturally plays an important role around the deconfinement
transition, but gluon quasiparticles dominate the thermodynamics at higher
temperatures. 

Given the role of the Polyakov loop as the order parameter for
the deconfinement phase transition in pure gauge theories, it is natural to
expect that the equilibrium free energy density $f$ can be extended to a
function $f(T,P)$ such that for any given temperature, $f(T,P)$ is minimized
at the equilibrium value of $P$. Here $P\,$\ is defined as a path-ordered
exponential in Euclidean time at finite temperature
\begin{equation}
P={\cal P}\exp ig\int_{0}^{\beta }dt\,\,A_{0} 
\end{equation}
where $\beta =1/T$. The deconfined phase of pure gauge theories is
associated with spontaneous breaking of a global symmetry under the center
of the gauge group. For $SU(3)$, the gauge theory is invariant under a
global symmetry transformation in which $P\rightarrow zP$,$\ $where \ \ $%
z\in \left\{ 1,e^{2\pi i/3},e^{4\pi i/3}\right\} $. It is convenient to
define $L=Tr_{F}\left( P\right) /N_{c}$, the normalized trace of $P$ in the
fundamental representation. If center symmetry is unbroken 
\begin{equation}
\left\langle L\right\rangle =\left\langle zL\right\rangle =z\left\langle
L\right\rangle 
\end{equation}
and $\left\langle L\right\rangle =0$. This is the confined, low temperature
phase of the gauge theory.
In the high temperature gluon plasma phase the symmetry is
spontaneously broken. In the spirit of mean field theory,
we will take $P$ to be a constant element of $SU(3)$,
ignoring spatial fluctuations. Thus gluon quasiparticles 
will be taken to move in
a constant gauge-invariant background.

In this letter, we study a broad class of models with free energies of the form 
\begin{equation}
f(T,P)=V(T,P)-p_{g}(T,P,M(T))+B(T,P,M(T)).
\label{eqn:broad_class}
\end{equation}
The first term $V(T,P)$ is phenomenological, and favors $L=0$ at low
temperature. This term is needed to induce a confined phase at low
temperatures, because the other two terms favor $L=1$ at all temperatures.
The second term is the negative of the quasiparticle
pressure $p_{g}$, and represents the
contribution of gluon quasiparticles moving in the presence of a
background Polyakov loop. This term is given by 
\begin{equation}
p_{g}(T,P,M(T))=-2T\int \frac{d^{3}k}{\left( 2\pi \right) ^{3}}Tr_{A}\,\ln
\left[ 1-e^{-\beta \omega _{k}}P\right] 
\end{equation}
where $Tr_{A}$ denotes the trace in the adjoint representation.
The pressure depends on the quasiparticle mass through the quasiparticle
energy $\omega_k$, assumed to obey the dispersion relation
$\omega _{k}^2=k^{2}+M^2\left( T\right)$, 
with a temperature-dependent mass $M(T)$.
The third term $B\left( T,P,M(T)\right) $ is the so-called bag term
\cite{Gorenstein:vm}.
It arises because the zero-quasiparticle state has a non-zero,
temperature-dependent energy $B$. If we write $B$ in such a way that it
depends on $T$ only through the quasiparticle mass $M$, then thermodynamic
consistency leads to an expression for $B\,$which may be written as 
\begin{equation}
B(T_{1},P,M\left( T_{1}\right) )-B(T_{0},P,M\left( T_{0}\right)
)=\int_{T_{0}}^{T_{1}}dT\,\,\,T\frac{dM^{2}}{dT}\int \frac{d^{3}k}{\left(
2\pi \right) ^{3}}\frac{1}{k^{2}}Tr_{A}\,\ln \left[ 1-e^{-\beta \omega
_{k}}P\right] 
\end{equation}
This term has no well-defined
connection to bag models of hadrons, although it plays
a superficially similar role in the free energy. After minimization of $%
f(T,P)$ with respect to $P$, the equilibrium pressure is given by the
negative of the minimum of the free energy density.

Previous work on quasiparticle models did not include Polyakov loop effects,
instead using various \textit{ad ho}c mechanisms to freeze out the
quasiparticle degrees of freedom as the temperature approaches $T_{c}$.
These models have $L=1$ at all temperatures, and are
thus incapable of reproducing the critical behavior of the deconfinement
transition seen in lattice simulations.

Modeling both the equation of state and 
the critical behavior is a much stronger
requirement than fitting the equation of state alone.
Fitting the lattice thermodynamic results exactly with a phenomenological
model is not difficult. \vspace{1pt}Any model with an adjustable parameter
which interpolates smoothly between $p=0\,$and the blackbody limit of $\pi
^{2}T^{4}/90$ per degree of freedom can be used to fit exactly the lattice
results for the pressure. 
Note that all other thermodynamic quantities are
obtainable from the pressure, so a perfect fit to the pressure in principle
gives a perfect fit to all. 
One procedure which exactly fits the lattice
pressure data is to determine $M(T)$ by requiring
that $p_{g}(T,P=I,M(T))$ equal the lattice result.
The mass so determined is shown in Figure 1.
\begin{figure}[htb]
\vspace{24pt}
\includegraphics[width=17pc]{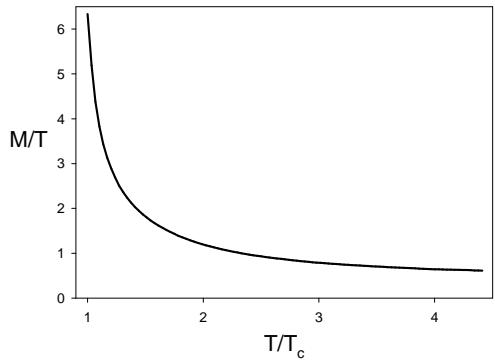}
\vspace{-60pt}
\caption{$M/T$ versus $T/T_c$ for exact fit to $p/T^4$.}
\label{fig:mexact}
\end{figure}
The rise in the quasiparticle mass near $T_{c}$ is typical of models which rely
heavily 
on the quasiparticle mass to suppress the pressure near $T_{c}$
\cite{Peshier:1995ty}.
Another approach is to use the function $p_{g}(T,P,M=0)$, and determine the
Polyakov loop from the lattice results. This leads to a Polyakov loop
behaving as shown in Figure 2. This behavior is similar to what is expected
for a first-order deconfining transition, and resembles recent results for
the renormalized Polyakov loop
\cite{Kaczmarek:2002mc,Dumitru:2003hp}
While either of these prescriptions fit
the lattice results perfectly, the two cannot be combined together without
additional assumptions. 
Nevertheless, these figures give an important
clue as to which effects are important at different temperatures. 
In the absence of Polyakov loop effects, a
temperature-dependent quasiparticle mass must show a steep increase as the
temperature is lowered towards $T_{c}$ in order to mimic the effects of the
deconfinement transition. Above about $2.5T_{c}$, $M(T)/T$ varies slowly
with temperature.
On the other hand, the model with a
temperature-dependent Polyakov loop shows a significant variation in $L$ in
the range $T_{c}-2.5T_{c}$, and $L$ appears to be nearly saturated at higher
temperatures. This suggests examining models where
Polyakov loop effects are most important in
the range $T_{c}-2.5T_{c}$, and the quasiparticle mass is important at
higher temperatures.

\begin{figure}[htb]
\vspace{42pt}
\includegraphics[width=17pc]{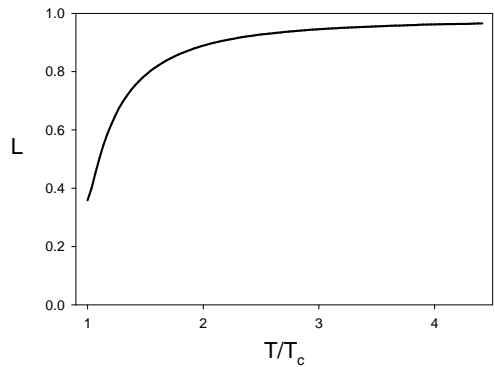}
\vspace{-60pt}
\caption{$L$ versus $T/T_c$ for exact fit to $p/T^4$.}
\label{fig:lexact}
\end{figure}

In previous work\cite{Meisinger:2001cq},
we have developed models for the $SU(N)$
deconfinement transition in pure gauge theories which fit within the general
class specified by equation (\ref{eqn:broad_class})
and give a good representation of the critical
behavior for all $N$. However, quasiparticle mass effects were not
considered: $M$ was identically zero, and hence $B$ was identically zero as
well. Here we study the effect of adding a temperature-dependent mass to
one of these models, with a potential $V$, of the form 
\begin{equation}
V(T,P)=-\frac{T}{R^{3}}\ln \left[ \mu \left( P\right) \right] +v_{0}.
\end{equation}
The function $\mu (P)$ is the Haar measure on the gauge group, 
a function of the
Polyakov loop which is maximized when $L=0$. With $V\,$
dominating the
free energy at low temperature, this term gives rise to the confined phase.
The parameter $R$ can be interpreted as the distance scale above which the
Polyakov loop enforces color neutrality. The limit $R\rightarrow \infty $
corresponds to global color neutrality only. In perturbation theory
only global color neutrality is enforced:
an infinite product of Haar measures is cancelled by ghost contributions in
the functional integral \cite{Gocksch:1993iy}.
In practice, we use the parameter $R\,$\
to set $T_{c}$ to the result of lattice simulations; 
we find in this model $R=1.38/T_{c}$.
The constant $v_{0}$ is used to match the pressure at one point on the
curve, which we have chosen to be $T_{c}$. In the absence of a
quasiparticle mass, this procedure fixes the only parameters of the model.
We show in
Figure 3 the dimensionless pressure \ \ $p/T^{4}\,$as a function of $T/T_{c}$
for this model ($M=0$), along with the results from reference
\cite{Boyd:1996bx}, which gives simulation data
extrapolated to the continuum limit.

\begin{figure}[htb]
\vspace{24pt}
\includegraphics[width=17pc]{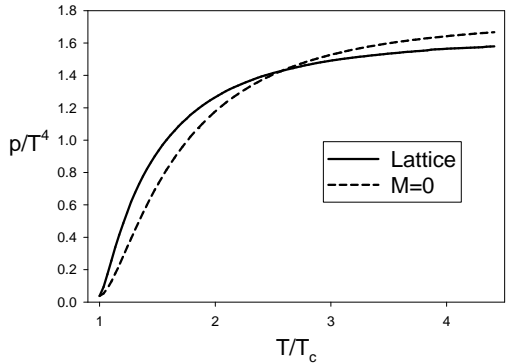}
\vspace{-60pt}
\caption{$p/T^4$ versus $T/T_c$ with $M=0$.}
\label{fig:predmodelb}
\end{figure}

We initially extend our model by taking $M(T)$ to be a linear function of
temperature: $M(T)=cT$. Because the effective mass depends on the temperature,
the so-called bag term must be included.
A least
squares fit to the dimensionless pressure gives $c=1.23$
and $R=1.61/T_{c}$.
Figure 4 shows $%
p/T^{4}$ for this model. The inclusion of the quasiparticle mass results in
a significant improvement in the goodness of fit over the model with $M\,\ $%
identically zero. The least-squares fit shows multiple nearby minima as a
function of the fitting parameter $c$; the fit yielding a global minimum is
shown.

\begin{figure}[htb]
\vspace{40pt}
\includegraphics[width=17pc]{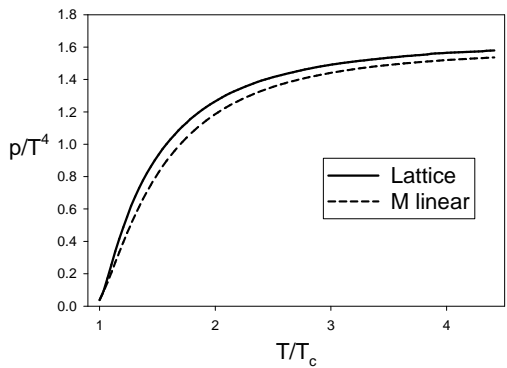}
\vspace{-60pt}
\caption{$p/T^4$ versus $T/T_c$ with $M(T)$ linear in $T$.}
\label{fig:predlinear}
\end{figure}

Previous quasiparticle models have taken $M(T)$ to be proportional to $g(T)T$%
, where $g(T)$ is the QCD coupling constant, running with 
temperature \cite{Peshier:1995ty,Levai:1997yx,Schneider:2001nf}.
We now consider a mass of the form $M(T)=g(T)T\sqrt{2}$, where $g(T)\,$is
given by 
\begin{equation}
g^{2}(T)=\frac{8\pi ^{2}}{11\ln \left( T/\Lambda \right) }
\end{equation}
as suggested by one-loop finite temperature perturbation theory
\cite{Peshier:1995ty,Kapusta:tk}.
\begin{figure}[htb]
\vspace{42pt}
\includegraphics[width=17pc]{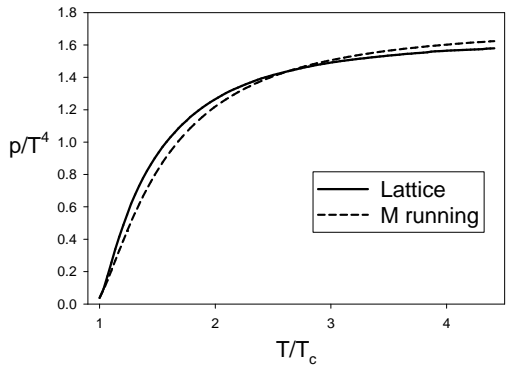}
\vspace{-60pt}
\caption{$p/T^4$ versus $T/T_c$ with $M(T)$ running.}
\label{fig:predrunning2}
\end{figure}
The sole adjustable parameter in fitting to simulation data is the
scale parameter $\Lambda $; as $\Lambda $ is varied, $R$ and $v_{0}$ are
adjusted to keep $T_{c}$ and $p(T_{c})$ fixed. A least-squares fit results
in $\Lambda =0.121T_{c}$, and $R=1.63/T_{c}$. The value of $\Lambda $ is
comparable to that found in 
\cite{Peshier:1995ty} and \cite{Levai:1997yx}. 
A direct comparison
is difficult, because both these works postulate different forms for the
running coupling in order to dramatically reduce quasiparticle effects at
low temperature. As may be seen from figure 5, the inclusion of running
coupling constant effects does not significantly improve the quality of the
fit to lattice simulations over the linear form. The quasiparticle mass is
shown as a function of temperature in figure 6; in both cases shown, the
mass is well-behaved near $T_{c}$. In figure 7, we show the dimensionless
interaction measure $\Delta/T^4$, where $\Delta =\varepsilon -3p$,
as a function of $%
T/T_{c}$ for the $M=0$ and $M=cT\,\ $fit compared with lattice results. For $%
\Delta $, the fit to a running mass is so close to the linear fit as to be
indistinguishable. Note that all these models correctly reproduce the
first-order character of the $SU(3)$ deconfinement transition, but the
combination of quasiparticle mass and Polyakov loop effects fit both the
latent heat discontinuity and the asymptotic tail. Our previous work
suggests that the peak in $\Delta $ is sensitive to the precise form of $V$
\cite{Meisinger:2001cq}.
On the other hand, the behavior of the order parameter $L$ is similar in
all the models considered here, resembling figure 2.

\begin{figure}[htb]
\vspace{42pt}
\includegraphics[width=17pc]{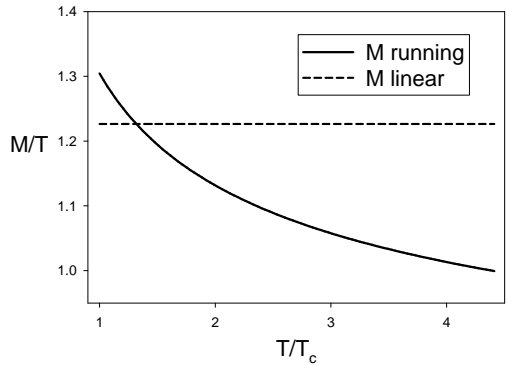}
\vspace{-60pt}
\caption{$M/T$ versus $T/T_c$ for $M(T)$ linear and running.}
\label{fig:mass}
\end{figure}
\vspace{-16pt}

\begin{figure}[htb]
\vspace{42pt}
\includegraphics[width=17pc]{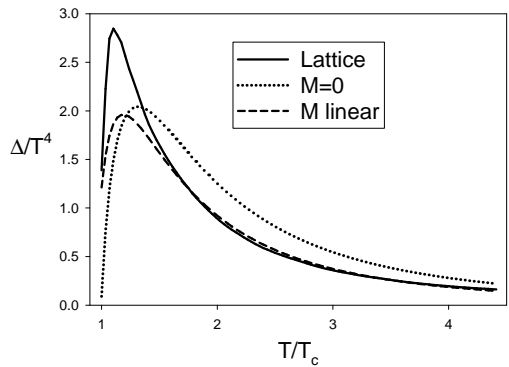}
\vspace{-60pt}
\caption{$\Delta/T^4$ versus $T/T_c$.}
\label{fig:dred}
\end{figure}

Within the class of models we have studied, neither a mass linear in
temperature nor a running quasiparticle mass \ by themselves give a good fit
to simulation results. We have studied the effect of a linear or running
quasiparticle mass alone by considering a a model with $V=0$, so that $L=1$
at all temperatures. The best fit to the lattice data with $V=0$ is worse
than the model with $M=0$, which we originally proposed in 
\cite{Meisinger:2001cq}.

The models we have studied here with $V\neq 0$ do a very
credible job of describing both the deconfining phase transition and
thermodynamic behavior in the deconfined phase. \vspace{1pt}It should be
apparent that many models will be able to fit lattice thermodynamic results
well, \ if a sufficient number of free parameters are used. The ability to
describe both the thermodynamic results and the deconfinement transition
with a small number of parameters, while not unique to the models described
here, seems very desirable. The introduction of $V$ in the free energy is
phenomenological, but based on our understanding of the deconfinement
transition. The onus of describing the temperature region just above $T_{c}$
is placed where it likely belongs, on the confinement mechanism. The
Polyakov loop $P\,$\ plays a central role in suppressing thermal
quasiparticle excitations at intermediate temperatures. The pressure is
naturally obtained by minimizing the free energy as a function of $P$.

We obtain a picture of the deconfined phase in which Polyakov loop effects
provide the dominant correction to blackbody thermodynamics in the
temperature regime from $T_{c}$ to approximately $2.5T_{c}$. At higher
temperatures, $L\simeq 1$, and quasiparticle mass effects can account for
most of the deviation from blackbody behavior. There is clear evidence from
higher temperatures that $M\varpropto T$, but the evidence for a one-loop
running coupling form is less compelling. Although we have not made a
systematic study, we believe that the behavior of the quasiparticle mass at
higher temperatures is not sensitive to the precise form of the confining
potential term $V$. On the other hand, a better choice for the form of $V$
could improve the fit of the interaction measure $\Delta $ to lattice
results just above $T_{c}$.

\vspace{1pt}There are several possible advances in our understanding of
finite temperature gauge theories
that would in turn lead to a better understanding of
the equation of state. The first, certainly the most fundamental and
probably the most difficult, is a better understanding of the mechanism of
confinement. An improved understanding would be reflected here as an
expression for $V$ \ based on fundamental physics rather than on
phenomenology. This is particulary important at intermediate temperatures
just above $T_{c}$. The second possibility is an analysis of lattice results
for the renormalized Polyakov loop in a form suitable for use in the
quasiparticle pressure term $p_{g}$
\cite{Kaczmarek:2002mc,Dumitru:2003hp}
A final possibility is a reliable
lattice determination of electric and magnetic screening masses. However,
screening masses are not synonymous with a gluon quasipartice masses, and
lattice results have been marred by ambiguities associated with lattice
gauge fixing.

The extension of these models by the
inclusion of quarks is directly relevant to experiment. However, our
inability to describe chiral symmetry breaking in a fundamental way forces
us to use couplings of Nambu-Jona Lasinio type to mimic QCD's chiral behavior.
Given that any model with a parameter that varies the pressure between
zero and the black-body result can give an exact match to simulation
results, the real possibility of reading too much out of the lattice data
increases as the number of fields and phenomenological parameters grows.
Nevertheless, a useful synthesis of our current understanding of
deconfinement and chiral symmetry restoration is likely possible.
The class of models which we have applied here to pure $SU(3)$
gauge theory also can 
be extended to other pure gauge theories as well,
with $SU(2)$ being most interesting because of its
second-order deconfining transition.

\end{document}